\documentclass[12pt]{iopart}
\eqnobysec
\usepackage{amssymb}
\usepackage{epsf}

\def\be{\begin{equation}}
\def\ee{\end{equation}}

\begin{document}
\jl{1}

\title{Gradient critical phenomena in the Ising quantum chain: surface behaviour}

\author{Mario Collura, Dragi Karevski\footnote[1]{Corresponding author: {\tt karevski@lpm.u-nancy.fr}} and Lo\"\i c Turban
}
\address{Groupe de Physique Statistique, D\'epartement Physique de la Mati\`ere et des Mat\'eriaux,
Institut Jean Lamour\footnote[2]{Laboratoire associ\'e au CNRS UMR 7198.}, CNRS---Nancy Universit\'e---UPV Metz,\\
BP 70239, F-54506 Vand\oe uvre l\`es Nancy Cedex, France
}

\begin{abstract}
We consider the influence of a power-law deviation from the critical coupling such that the system is critical at its surface. We develop a scaling theory showing that such a perturbation introduces a new length scale which governs the scaling behaviour of the density profiles as well as the finite-size behaviour of the surface properties. Exact results are obtained for the Ising quantum chain when the perturbation varies linearly whereas the quadratic perturbation is mainly studied numerically. The scaling theory is well confirmed in both cases. 
\end{abstract}

\section{Introduction}
Real systems are in general inhomogeneous due to the presence of impurities, line defects, boundaries, aperiodicity or disorder. Such inhomogeneties may have a strong influence on the critical properties in the vicinity of a second order phase transition \cite{igloi93,diehl86,grimm97,mccoy72,berche04}.
Depending on the relevance of the perturbation introduced by the inhomogeneity, the universality class governing the critical behaviour may change; in some cases, the critical singularities may even be suppressed.

Besides this, the presence of spatially varying external fields (magnetic, gravitational or thermal) also influences the critical properties of a system. 
The effect of gravity on phase coexistence was studied in  \cite{rogiers93,carlon97,carlon98}. Phase separation induced by a thermal gradient was considered in \cite{platten93,assenheimer94,kumaki96}. The application of a gradient perturbation was used as a tool for high-precision estimates of percolation threshold and exponents in \cite{sapoval85,rosso85,rosso86,ziff86}.

In general the effect of a spatially varying field on a critical system is to smooth out the critical singularities. The reason for this is that the perturbation generated by the inhomogeneous field  leads to a departure from the critical point which 
introduces a finite length scale in the problem \cite{sapoval85}. One may think about the usual finite-size scaling theory as a particular example where the spatially varying field is a confining potential of a box-like type \cite{binder83}. 

For a field generating  a power-law deviation  from the critical point, $\Delta(z)=g z^\omega$, the typical length scale introduced around the critical locus, $z=0$, is $\ell\sim |g|^{-\nu/(1+\nu\omega)}$ where $\nu$ is the correlation exponent.
Far away from this critical region, for $|z|\gg \ell$, the correlation length is small compared to the scale of the spatial change and the local properties at $z$ are determined by the corresponding homogeneous system with a constant deviation to the critical point given by the value $\Delta(z)$.  Based on these observations, a scaling theory for the density profiles in the presence of gradient-field inhomogeneities was developed in \cite{platini07}. 
The validity of the scaling theory was checked  at the mean-field level within Ginsburg-Landau theory and by comparing its predictions to the exact solution of an Ising quantum chain in a linearly varying transverse field. 

The same scaling predictions were tested for the Ising quantum chain with a linear-to-flat profile of the transverse field \cite{zurek08}, for a spin-1 ferromagnetic Bose-Einstein condensate in a space-varying magnetic field \cite{damski09} and for a classical Ising lattice gas confined in a two-dimensional harmonic trap \cite{campostrini09}.
The entanglement entropy for Ising and XX quantum chains with linearly varying fields or couplings was studied in \cite{eisler09}. Quite recently the trap-size scaling was examined for the XY and Bose-Hubbard quantum chains \cite{campostrini09b}.

One may mention a formal analogy noticed in \cite{zurek08} between the spatial inhomogeneous case discussed here and the quench dynamics situation where a system is driven through a critical point at finite rate. During the quench, far away from the critical point, the system is able to adjust its state adiabatically as far as its relaxation time is much shorter than the typical time associated with the quench. On approaching the critical point, due to the critical slowing down, the typical quench time becomes much shorter than the system relaxation time and the evolution switch approximately to an impulse regime. Close to the critical point, the typical time scale during which the state is frozen  (impulse evolution) is given by equating the system relaxation time (inverse of the first gap for a closed quantum system) to the typical quench time scale. In the quantum case this leads to
the self consistent equation $\tau=|1/\epsilon(\tau)|^{\nu z}$ where $\epsilon(t)$ is the deviation to the critical point at time $t$, $z$ the dynamical exponent and $\nu$ the correlation length exponent. For a power law deviation $\epsilon(t)=g t^\omega$ one obtains $\tau \sim g^{-z\nu/(1+\omega z\nu)}$ in complete analogy with the spatial case. 
Drawing on this analogy, we see that to the adiabatic evolution regime corresponds a ``super-static'' regime where, locally,  the physical properties are determined solely invoking a local equilibrium hypothesis. On the contrary, on approaching (either in time or in space) the critical locus, the local equilibrium hypothesis  (time- or space-like)
breaks down due to the divergence of the typical time (relaxation) or space scale (correlation length).

All these studies were made in reference to the bulk critical behaviour, i.e., when the locus of the critical field value is deep into the bulk of the system. 
However, if one shifts the critical locus toward a free surface, when close enough to it, the scaling behaviour of local quantities should depend on the surface critical properties. In the present work  we consider such a situation for a semi-infinite system with an ordinary surface transition. The paper is organized in the following way:
in the next section we present the scaling arguments 
for a semi-infinite system with a power-law deviation from the critical point when the critical locus is at the surface.
The third section is devoted to the semi-infinite Ising chain in a linearly or quadratically varying transverse field. In the linear case we solve analytically the problem while in the quadratic case the results are obtained through an exact numerical diagonalization. In both cases our findings are in perfect agreement with the scaling predictions.  Finally we summarize our results and indicate some possible developments in the last section.

\section{Scaling arguments}
In the next sections we study either linear (gradient) perturbations, for which exact results can be obtained, or quadratic ones. Thus here we shall examine the scaling behaviour for an arbitrary power law variation of the couplings. 

The system is semi-infinite, occupies the half-space $z>0$ with free boundary conditions at $z=0$. The perturbed coupling deviates from its critical value $K_{\rm c}$ as\footnote{For $g<0$ equation \eref{2-1} leads to negative values of the coupling for sufficiently large $z$. The value of $g$ and the size of the system should be chosen in order to avoid the occurence of negative values.}
\be
K(z)-K_{\rm c}=\Delta(z)=gz^\omega\,, \qquad \omega>0
\label{2-1}
\ee
Thus the coupling is critical at the surface $z=0$. The system is ordered (disordered) in the bulk when $g>0$ ($g<0$). The perturbation introduces a crossover length $\ell$ giving the width of the surface region which diverges when $g$ vanishes. With a decaying perturbation ($\omega<0$) the discussion of the scaling behaviour is similar but the physics is quite different since {\it the bulk remains critical} while the surface critical behaviour is mofified when the perturbation is relevant ($0>\omega>-1/\nu$) or marginal ($\omega=-1/\nu$)\cite{hilhorst81,cordery82,burkhardt82a,burkhardt82b,bloete83,bloete85}.

Changing the length scale by a factor $b$, the thermal perturbation $\Delta(z)$ with scaling dimension $y_t=1/\nu$ where $\nu$ is the  correlation length  exponent, transforms 
as\footnote{The following discussion extends to other types of perturbation provided the appropriate scaling dimension replaces $y_t$.}
\be
g'{z'}^\omega=b^{1/\nu}gz^\omega=g'\left(\frac{z}{b}\right)^\omega\,,
\label{2-2}
\ee
so that 
\be
g'=b^{\omega+1/\nu}g\,.
\label{2-3}
\ee
As mentioned above $g$ grows under rescaling and the perturbation is relevant when $\omega>-1/\nu$.

The crossover length transforms as 
\be 
{\ell}'=\frac{\ell}{b}=\ell(g')=\ell\left(b^{\omega+1/\nu}g\right)\,,
\label{2-4}
\ee
with $b=|g|^{-\nu/(1+\nu\omega)}$, one finally obtains
\be
\ell=A|g|^{-\nu/(1+\nu\omega)}\,,
\label{2-5}
\ee
for the crossover length introduced by the thermal  perturbation. The result of equation (2.5) in \cite{platini07} is recovered when $\omega=1$.

As in \cite{platini07} this result can be obtained self-consistently~\cite{sapoval85,eisler09} by relating $\ell$ to the local value of the correlation length $\xi(\ell)$ so that
\be
\ell\propto \xi(\ell)\propto |\Delta(l)|^{-\nu}\propto (|g|\ell^\omega)^{-\nu}\,,
\label{2-6}
\ee
which gives back~\eref{2-5} when solved for $\ell$.
 
Let us nown examine the scaling behaviour of the profile $\varphi(z,g)$ associated with some bulk operator $\varphi$ with scaling dimension $x_\varphi$. This may be the order parameter $m$ or the singular part of the energy density $e$. It transforms as
\be
{\varphi}'=\varphi(z',g')=b^{x_\varphi}\varphi(z,g)\,,
\label{2-7}
\ee
so that, using \eref{2-3}
\be
\varphi(z,g)=b^{-x_\varphi}\varphi\!\left(\frac{z}{b},b^{\omega+1/\nu}g\right)\,.
\label{2-8}
\ee
With $b=|g|^{-\nu/(1+\nu\omega)}\propto\ell$ one obtains
\be
\varphi(z,g)=|g|^{\nu x_\varphi/(1+\nu\omega)}\Phi\!\left(|g|^{\nu/(1+\nu\omega)} z\right)
\propto\ell^{-x_\varphi}\Phi\! \left(\!A\,\frac{z}{\ell}\right)\,,
\label{2-9}
\ee
where the last expression follows from \eref{2-5}. The prefactor exhibits the 
finite-size behaviour expected for a critical system with size $\ell$.
In the same way, for a surface operator $\varphi_{\rm s}$ with scaling dimension $x_{\varphi_{\rm s}}$ at the ordinary surface transition, in a system with size $L$, one obtains:
\be
\varphi_{\rm s}(L,g)=|g|^{\nu x_{\varphi_{\rm s}}/(1+\nu\omega)}\Phi_{\rm s}\! \left(|g|^{\nu/(1+\nu\omega)} L\right)
\propto\ell^{-x_{\varphi_{\rm s}}}\Phi_{\rm s}\! \left(\!A\,\frac{L}{\ell}\right)\,.
\label{2-10}
\ee
When $L\ll\ell$, the effect of the perturbation is negligible and the surface operator displays the usual finite-size behaviour for a critical system $\varphi_{\rm s}(L,g)\sim L^{-x_{\varphi_{\rm s}}}$. As a consequence, the scaling function behaves as:
\be
\Phi_{\rm s}(u)\sim u^{-x_{\varphi_{\rm s}}}\,,\qquad u\ll 1\,.
\label{2-11}
\ee
When $L/\ell\to\infty$, $\varphi_{\rm s}$ no longer depends on $L$, so that:
\be\fl
\lim_{u\to\infty} \Phi_{\rm s}(u)={\rm const}\,,\qquad \lim_{L/\ell\to\infty}\varphi_{\rm s}(L,g)\propto\ell^{-x_{\varphi_{\rm s}}}={\rm const}\,|g|^{\nu x_{\varphi_{\rm s}}/(1+\nu\omega)}\,.
\label{2-12}
\ee
Such a behaviour is expected with $e_{\rm s}$ for any sign of $g$ and with $m_{\rm s}$ for $g>0$.

In the surface region $z\ll\ell$ the prolile in \eref{2-9}, when non-vanishing, should scale as $\ell^{-x_{\varphi_{\rm s}}}$ 
\be
\varphi(z,g)\propto\ell^{-x_\varphi}\Phi\! \left(\!A\,\frac{z}{\ell}\right)\propto\ell^{-x_{\varphi_{\rm s}}}\,,\qquad \frac{z}{\ell}\ll 1\,,
\label{2-13}
\ee
leading to:
\be
\Phi(u)\sim u^{x_{\varphi_{\rm s}}-x_\varphi}\,,\qquad u\ll 1\,.
\label{2-14}
\ee

The order parameter $m(z,g)$, with scaling dimension $x_{\rm m}=\beta/\nu$,  is non-vanishing in the ordered phase $g>0$. Then, outside the surface region when $z\gg\ell$, one expects the same local critical behaviour as for the homogeneous system with the same value of the coupling, i.e.,
\be
m(z,g)\propto\Delta(z)^{\beta}\propto\left(gz^\omega\right)^{\nu x_{\rm m}}\,,\qquad g>0\,.
\label{2-15}
\ee
for not too large values of $\Delta(z)$. Comparing~\eref{2-15} to the first expression of the profile in \eref{2-9} with $\varphi=m$, one obtains the asymptotic behaviour of the scaling function:
\be
\Phi_{\rm m}(u) \sim u^{\omega\nu x_{\rm m}}\,,\qquad u\gg 1\,,\qquad g>0\,.
\label{2-16}
\ee

\section{Quantum Ising chain in a varying transverse field}

\subsection{Quantum Hamiltonian}
We consider the quantum Ising chain in a varying transverse field with Hamiltonian \cite{pfeuty70}
\be
{\cal H}=-\frac{1}{2}\sum_{l=1}^{L-1}\sigma^x_l\sigma^x_{l+1}
-\frac{1}{2}\sum_{l=1}^L h_l\,\sigma^z_l\,, \qquad h_l=h(1-gl^\omega)\,,
\label{3-1}
\ee
where the $\sigma^{x,z}$ are Pauli spin operators. This Hamiltonian governs the extreme anisotropic limit of the classical two-dimensional Ising model on the square lattice \cite{suzuki71,fradkin78,kogut79}. As explained in \cite{platini07}, when the classical system has horizontal couplings $K_2$ and varying dual vertical couplings $K_1^*(l)=K_1^*(1-gl^\omega)$, the row-to-row transfer operator $\cal T$ takes the form
\be
{\cal T}=1-2K_2{\cal H}\,.
\label{3-2}
\ee
when $K_1^*\to 0$, $K_2\to 0$ while $h=K_1^*/K_2$ remains constant. Universality holds in this limit and the critical properties of the quantum chain are the same as those of the two-dimensional classical system with \cite{igloi93}:
\be\fl
\nu=1\,,\quad  x_{\rm e}=1\,,\quad  x_{\rm m}=1/8\,,\quad x_{\rm e_s}=2\,,
\quad x_{\rm m_s}=1/2\ {\rm (ordinary\ transition)}\,.
\label{3-2b}
\ee

In the following we take $h=1$ and the quantum chain is truly critical when $g=0$ and $L\to\infty$.

\subsection{Diagonalization}
Under a Jordan-Wigner transformation \cite{jordan28} the Hamiltonian \eref{3-1} is changed into a quadratic form in fermion creation and destruction operators which is diagonalized through a Bogoliubov transformation \cite{lieb61,kogut79}. Then 
\be 
{\cal H}= \sum_{q=0}^{L}\varepsilon_{q}\left(\eta_{q}^{\dag}\eta_{q}-\frac{1}{2}\right)\,,
\label{3-3}
\ee
where $\eta_{q}^{\dag}$  ($\eta_{q}$ ) are diagonal fermion creation
(anihilation) operators. The fermionic excitation energies $\varepsilon_q$  are obtained through the solution of one of the following eigenvalue equations
\begin{eqnarray}
&&h_{l-1}\,\phi_q(l-1)+\left[\varepsilon_q^2-1-h_l^2\right]\phi_q(l)+h_{l}\,\phi_q(l+1)=0\,,
 \nonumber\\
&&h_{l}\,\psi_q(l-1)+\left[\varepsilon_q^2-1-h_l^2\right]\psi_q(l)+h_{l+1}\,\psi_q(l+1)=0\,
\label{3-4}
\end{eqnarray}
with the boundary conditions
\begin{eqnarray}
&&\phi_q(0)-\phi_q(1)=0\,,\qquad \phi_q(L+1)=0\,,\nonumber\\
&&\psi_q(0)=0\,,\qquad \psi_q(L)-\psi_q(L+1)=0\,.
\label{3-5}
\end{eqnarray}

In these equations $\psi_q(l)$ and $\phi_q(l)$ are the normalized eigenvectors.

The scaling variable of the quantum chain with $\nu=1$ is
\be
u=|g|^{\nu/(1+\nu\omega)}l=|g|^{1/(\omega+1)}l\,.
\label{3-6}
\ee
In the scaling limit $g\to 0$, $L\to\infty$ while keeping $\theta=gL^\omega$ constant, we have
$u(l+1)-u(l)=|g|^{1/(\omega+1)}\to 0$ and $u(L)=|g|^{-1/[\omega(\omega+1)]}\theta^{1/\omega}\to\infty$. Thus the system is semi-infinite in $u$ which is a continuous variable. The finite-difference equations in \eref{3-4} can be rewritten as differential equations in the reduced variable $u$. Keeping the leading contributions in $g$, of order $|g|^{2/(\omega+1)}$, one obtains the following differential equations:
\begin{eqnarray}
\fl &&\frac{{\rm d}^2\phi}{{\rm d} u^2}+\left[\left(\frac{\varepsilon}{|g|^{1/(\omega+1)}}\right)^2 
+{\rm sign}(g)\,\omega u^{\omega-1}-u^{2\omega}\right]\phi(u)=0\,,
\quad 
\left.\frac{{\rm d}\phi}{{\rm d} u}\right|_{u=0}\!\!\!\!\!\!=0\,,
\quad \phi(\infty)=0\,,
\nonumber\\
\fl &&\frac{{\rm d}^2\psi}{{\rm d}u^2}
+\left[\left(\frac{\varepsilon}{|g|^{1/(\omega+1)}}\right)^2
-{\rm sign}(g)\,\omega u^{\omega-1}-u^{2\omega}\right]\psi(u)=0\,,
\quad \psi(0)=0\,,
\quad 
\left.\frac{{\rm d}\psi}{{\mathrm d} u}\right|_{u\to\infty}\!\!\!\!\!\!\!\!\!=0\,.
\label{3-7}
\end{eqnarray}
When $g$ changes sign the two equations are exchanged but the boundary conditions remain the same. 
\begin{figure} [tbh]
\epsfxsize=7cm
\begin{center}
\mbox{\epsfbox{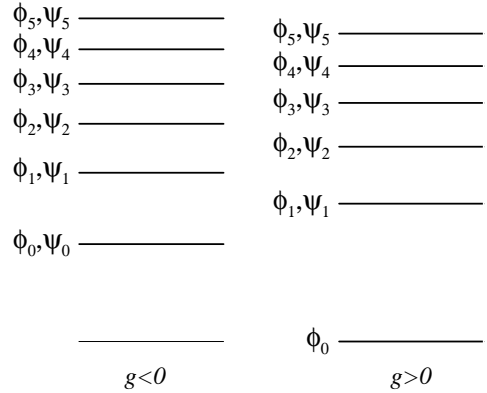}}
\end{center} 
\vskip -.8cm
\caption{Excitation spectra of the quantum spin chain. When $g>0$ (ordered phase) the lowest excitation vanishes and the ground state is degenerate.}
\label{fig1}  \vskip 0cm
\end{figure}

When $\omega=1$ we have:
\begin{eqnarray}
&&\frac{{\rm d}^2\phi}{{\rm d} u^2}+\left[\frac{\varepsilon^2}{|g|}
+{\rm sign}(g)-u^2\right]\phi(u)=0\,,
\quad 
\left.\frac{{\rm d}\phi}{{\rm d} u}\right|_{u=0}\!\!\!\!\!\!=0\,,
\quad \phi(\infty)=0\,,
\nonumber\\
&&\frac{{\rm d}^2\psi}{{\rm d}u^2}
+\left[\frac{\varepsilon^2}{|g|}
-{\rm sign}(g)-u^2\right]\psi(u)=0\,,
\quad \psi(0)=0\,,
\quad 
\left.\frac{{\rm d}\psi}{{\mathrm d} u}\right|_{u\to\infty}\!\!\!\!\!\!\!\!\!=0\,,
\label{3-8}
\end{eqnarray}
with $u=g^{1/2}z$ where $z\geq 0$ is the continuous coordinate along the quantum chain in the scaling limit.

These equations are eigenvalue equations of the harmonic oscillator form\footnote{For an oscillator with mass $M$, angular frequency $\Omega$, such that $M\Omega/\hbar=1$.}
\be
\frac{{\rm d}^2\chi_n}{{\rm d} u^2}+(2n+1-u^2)\chi_n(u)=0\,,
\label{3-9}
\ee
with eigenfunctions $\chi_n(g^{1/2}z)$, vanishing at infinity and normalized on $z\geq 0$, which are given by
\be\fl
\chi_n(u)=\sqrt{\frac{2}{2^n n!}}\left(\frac{g}{\pi}\right)^{1/4}\e^{-u^2/2}H_n(u)\,,
\qquad H_n(u)=(-1)^n \,\e^{u^2}\frac{{\rm d}^n}{{\rm d} u^n}\,\e^{-u^2},
\label{3-10}
\ee
where $H_n(u)$ is the Hermite polynomial of order $n$.
In \eref{3-8}, at $u=0$, the Neumann boundary condition for $\phi$ selects even values of $n$
whereas the Dirichlet boundary conditon for $\psi$ selects odd values of $n$.
Thus for the {\it ordered system} ($g>0$) one obtains:
\begin{eqnarray}
\fl \varepsilon_p&=&\sqrt{4pg}\,,\qquad \phi_p(u)=\chi_{2p}(u)\,,\qquad \psi_p(u)=\chi_{2p-1}(u)\,,\qquad p=1,2,3,\cdots\nonumber\\
\fl \varepsilon_0&=&0\,,\qquad \phi_0(u)=\chi_0(u)\,,\qquad \psi_0(u)=0\,.
\label{3-11}
\end{eqnarray}
The vanishing excitation $\epsilon_0$ is a consequence of the twofold degeneracy of the ordered ground-state. The behaviour is different for the {\it disordered system} ($g<0$) with a  non-degenerate ground state. Then the lowest excitation is non-vanishing:
\be\fl
\varepsilon_p=\sqrt{2(2p+1)|g|}\,,\quad \phi_p(u)=\chi_{2p}(u)\,,\quad \psi_p(u)=\chi_{2p+1}(u)\,,\quad p=0,1,2,\cdots
\label{3-12}
\ee
The lowest part of the excitation spectra is shown in figure \ref{fig1}.

\begin{figure} [tbh]
\epsfxsize=14cm
\hskip 16mm
\mbox{\epsfbox{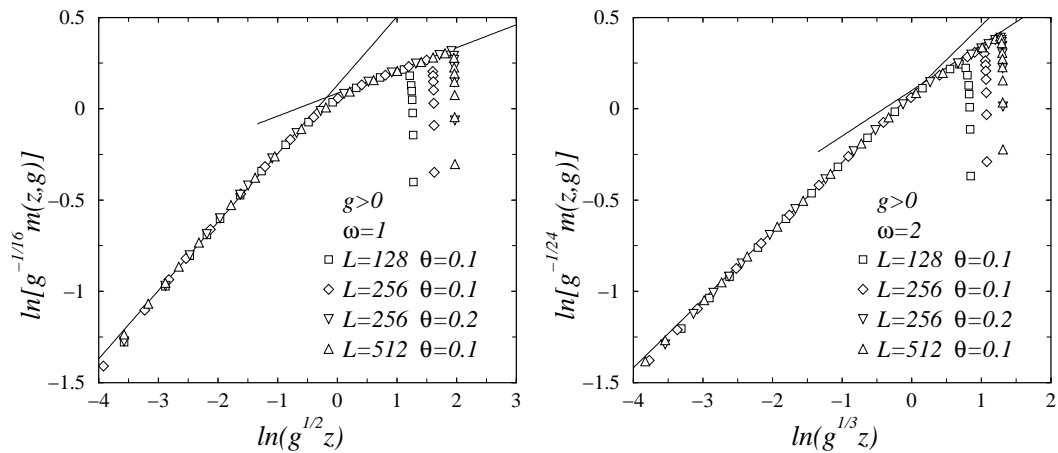}}
\vskip -.2cm
\caption{Scaling of the magnetization profile for  $\omega=1$ (left) and $\omega=2$ (right). the lines indicate the slopes expected from \eref{2-14} for $z\ll\ell$ and \eref{2-16} for $z\gg\ell$.}
\label{fig2}  \vskip 0cm
\end{figure}

\subsection{Magnetization profile and surface magnetization}
When $g>0$ the Z$_2$ symmetry is spontaneously broken in the thermodynamic limit. In a finite system it can be broken by fixing the last spin ($\sigma^x_L=\pm 1$ when $h_L=0$). Then the magnetization is non-vanishing. At $l1$, it is given by the off-diagonal matrix element $|\langle \sigma|\sigma_l^x|0\rangle|$ where $|0\rangle$ is the ground-state of $\cal H$ and $|\sigma\rangle =\eta_0^\dag|0\rangle$ is the lowest excited state with one fermion \cite{platini07}. The matrix element is evaluated by expanding $\sigma_l^x$ in diagonal fermionic operators. 

The magnetization profile $m(l)$ can be expressed as a determinant \cite{berche96}
\be
m(l)=\left| \begin{array}{lllll}
H_{1}&G_{1,1}&G_{1,2}& \dots&G_{1,\,l-1}\\
H_{2}&G_{2,1}&G_{2,2}& \dots&G_{2,\,l-1}\\
\vdots&\vdots&\vdots&      &\vdots\\
H_{l-1}&G_{l-1,1}&G_{l-1,2}& \dots&G_{l-1,\,l-1}\\
H_l&G_{l,1}&G_{l,2}& \dots&G_{l,\,l-1}\end{array}
\right|
\label{3-13}
\ee
where
\be
H_{j}=\phi_{0}(j)\,,\qquad G_{j,k}=-\sum_{n=0}^L\phi_n(j)\psi_n(k)\,.
\label{3-14}
\ee

The rescaled magnetization profiles are shown in figure \ref{fig2} where a good data collapse is obtained. The slopes are in good agreement with the values following from \eref{2-14} and \eref{2-16}:  $x_{\rm m_s}-x_{\rm m}=3/8$ for $z\ll\ell$, $\omega\nu x_{\rm m}=1/8$ when $\omega=1$ and $1/4$  when $\omega=2$ for $z\gg\ell$.

\begin{figure} [tbh]
\epsfxsize=14cm
\hskip 17mm
\mbox{\epsfbox{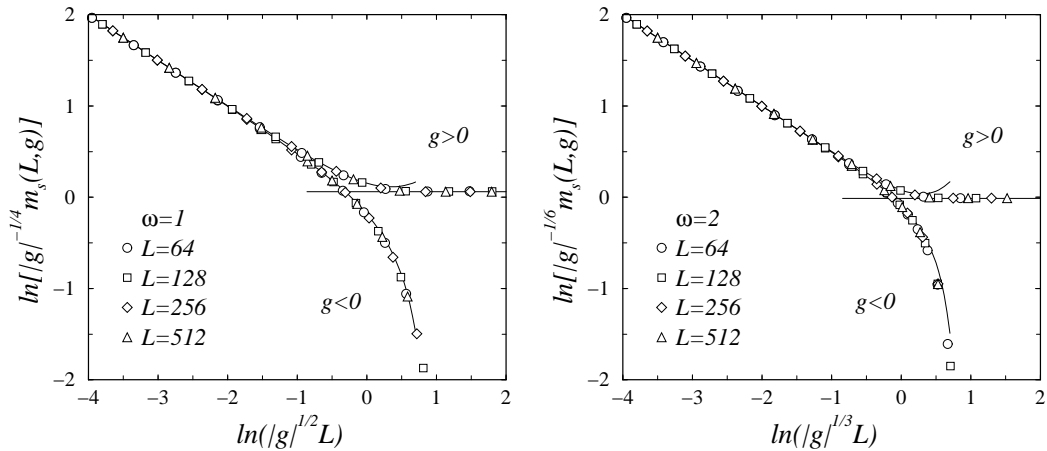}}
\vskip -.2cm
\caption{Finite-size scaling function of the surface magnetization for $\omega=1$ (left) and $\omega=2$ (right). The lines indicate the limiting behaviours given in \eref{3-17} and \eref{3-18}.}
\label{fig3}  
\vskip 0cm
\end{figure}
\begin{figure} [tbh]
\epsfxsize=7cm
\begin{center}
\mbox{\epsfbox{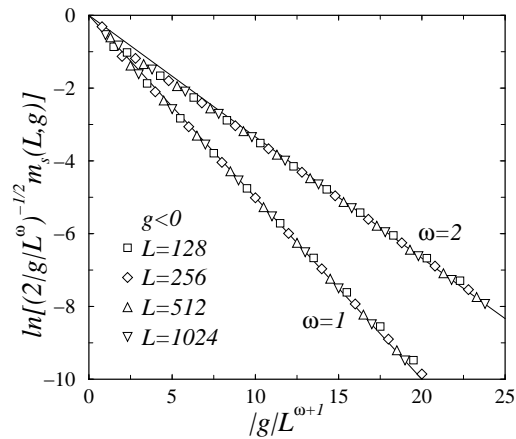}}
\end{center} 
\vskip -.8cm
\caption{Scaling function of the surface magnetization, reduced to its exponential part. In the semi-logarithmic scale, the decay is linear in $|g|L^{\omega+1}$ in agreement with the result of \eref{3-18} for $g<0$ (solid lines).}
\label{fig4}  \vskip 0cm
\end{figure}

When $l=1$, the determinant $m(1)$ reduces to the normalized eigenvector $\phi_0(1)$ which can be evaluated for arbitrary values of the couplings \cite{peschel84,karevski00}. It gives the surface magnetization:
\be
m_{\rm s}=\phi_0(1)=\left[1+\sum_{k=1}^{L-1}\prod_{l=1}^{k}h_l^2\right]^{-1/2}\,.
\label{3-15}
\ee
In the unperturbed system this leads to $m_{\rm s}\sim L^{-1/2}$ and $x_{\rm m_s}=1/2$ as indicated in \eref{3-2b}.

In the scaling limit ($L\to\infty$, $\ell\to\infty$, $u(z)=Az/\ell$) the surface magnetization is given by $\phi_0(u)$ at $u(1)=0$ so that, according to \eref{3-10} and \eref{3-11}, when $\omega=1$:
\be
m_{\rm s}=\sqrt{2}\left(\frac{g}{\pi}\right)^{1/4}\,.
\label{3-16}
\ee

The general case is discussed in the appendix where the finite-size behaviour of the surface magnetization in \eref{3-15} is studied in different limiting cases. When the scaling variable $u(L)=|g|^{1/(\omega+1)}L\ll1$ the leading behaviour 
is that of a finite-size critical system with
\be
m_{\rm s}\simeq L^{-1/2}\left[1+{\rm sign}(g)
\frac{\left(|g|^{1/(\omega+1)}L\right)^{\omega+1}}{(\omega+1)(\omega+2)}\right]\,,
\label{3-17}
\ee
in agreement with~\eref{2-10} and~\eref{2-11}. 

When $u(L)\gg1$ and $L$ such that $|g|L^\omega=u^{\omega+1)}/L\ll1$, one obtains
\begin{eqnarray}
m_{\rm s}&\simeq& A_\omega g^{1/[2(\omega+1]}\,,\qquad A_m(\omega)=\frac{\left[2(\omega+1)^\omega\right]^{1/[2(\omega+1)]}}
{\sqrt{\Gamma\left(\frac{1}{\omega+1}\right)}}\,.\qquad g>0\,.\nonumber\\
m_{\rm s}&\simeq& \sqrt{2|g|L^\omega}\,\exp\left(-\frac{|g|L^{\omega+1}}{\omega+1}\right)\,,
\qquad g<0\,.
\label{3-18}
\end{eqnarray}
The result for $m_{\rm s}$ in the ordered phase agrees with \eref{3-16} when $\omega=1$ and its $g$-dependence is the one expected from \eref{2-12}.

Numerical data  for the finite-size scaling function of the surface magnetization are compared to analytical results in figure~\ref{fig3}. The linear decay for $L\ll\ell$ has a slope equal to $-x_{\rm m_s}=-1/2$ which follows from \eref{2-11} and \eref{3-17}. The constant values for $L\gg\ell$, given by \eref{3-18} for $g>0$, are respectively $\ln[A_m(1)]=0.06039\dots$  for $\omega=1$ and $\ln[A_m(2)]=-0.01098\dots$ for $\omega=2$. The exponential decay for $g<0$ and $L\gg\ell$ is compared to the asymptotic analytical result of equation \eref{3-19} in figure \ref{fig4}.

\subsection{Energy density profile and surface energy density}

\begin{figure} [tbh]
\epsfxsize=14.5cm
\hskip 12mm\mbox{\epsfbox{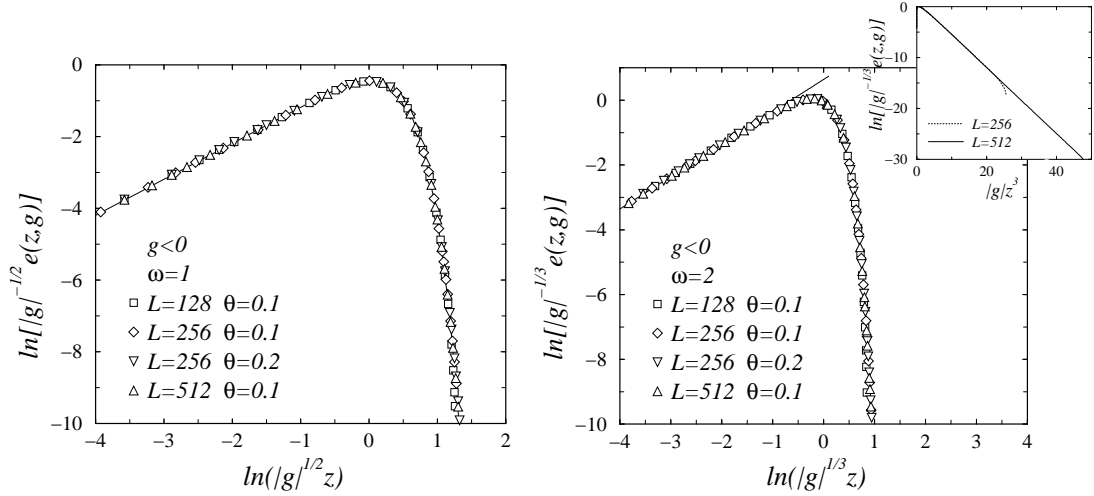}}
\vskip -.2cm
\caption{Scaling function of the energy density profile in the disordered phase. For $\omega=1$ (left) the solid line shows the  profile obtained in \eref{3-20} in the scaling limit. For $\omega=2$ (right) the solid line indicates the slope expected from \eref{2-14} when $z\ll\ell$ and the inset shows the exponential decay in the variable $|g|z^3$.}
\label{fig5}  \vskip 0cm
\end{figure}

\begin{figure} [tbh]
\epsfxsize=14.5cm
\hskip 12mm\mbox{\epsfbox{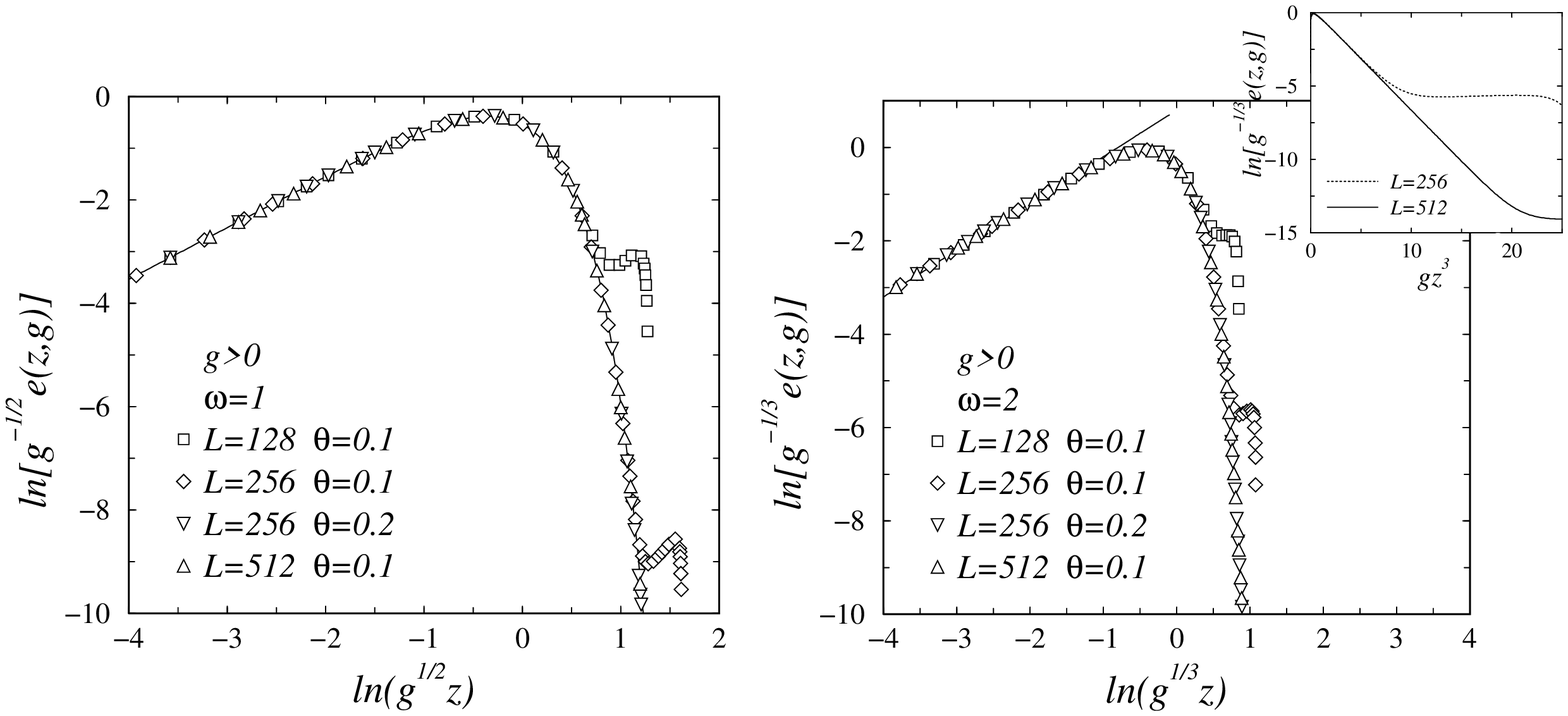}}
\vskip -.2cm
\caption{Scaling function of the energy density profile in the ordered phase. For $\omega=1$ (left) the solid line shows the  profile obtained in \eref{3-21} in the scaling limit. For $\omega=2$ (right) the solid line indicates the slope expected from \eref{2-14} when $z\ll\ell$ and the inset shows the exponential decay in the variable $|g|z^3$. The finite-size anomalies near the second surface at $z=L$ are discussed in the text.}
\label{fig6}  \vskip 0cm
\end{figure}

The scaling behaviour of the energy density can be studied on the off-diagonal matrix element $|\langle \epsilon|\sigma_l^z|0\rangle|$ where  $|\epsilon\rangle =\eta_1^\dag\eta_0^\dag|0\rangle$ is the lowest excited state with two fermionic excitations (see \cite{platini07} for details). The expansion  of $\sigma_l^z$ in terms of diagonal fermions leads to~\cite{berche90}
\be
e(l)=|\psi_1(l)\phi_0(l)-\psi_0(l)\phi_1(l)|\,.
\label{3-19}
\ee

When $\omega=1$, in the scaling limit with $u=u(z)=|g|^{1/2}z$, equations \eref{3-12}, \eref{3-13} and \eref{3-14} can be used to calculate the energy profile which is given by
\begin{eqnarray}
e(z)&=&|\chi_3(u)\chi_0(u)-\chi_1(u)\chi_2(u)|
=A_e^-|g|z\left(1+\frac{2|g|z^2}{\sqrt{3}}\right)\e^{-|g|z^2}\,,\nonumber\\
A_e^-&=&\frac{2(\sqrt{3}-1)}{\sqrt{\pi}}\,,\qquad g<0\,,
\label{3-20}
\end{eqnarray}
when the system is disordered and by
\be
e(z)=|\chi_1(u)\chi_0(u)|
=A_e^+gz\e^{-|g|z^2}\,,\qquad
A_e^+=2\sqrt{\frac{2}{\pi}}\,,\qquad g>0\,,
\label{3-21}
\ee
in the ordered phase.

The numerical data for the scaled energy profiles are shown in figure \ref{fig5} for $g<0$ and figure \ref{fig6} for $g>0$. A goog data collapse is obtained except for a finite-size anomaly near $L$ when $g>0$ which is sent to infinity in the scaling limit. 
When $\omega=1$ the convergence to the exact profiles obtained 
in the scaling limit is rapid. In the log-log scale the slope of the linear growth for $z\ll\ell$ is equal to $x_{\rm e_s}-x_{\rm e}=1$ as expected from \eref{2-14} and obtained analytically in \eref{3-20} and \eref{3-21}. For $z\gg\ell$ the decay is exponential in the variable $|g|z^2$. The insets show that for $\omega=2$ the scaled profiles display a similar decay in the variable $|g|z^3$. More generally one excepts an exponential decay in the variable $|g|z^{\omega+1/\nu}$ according to \eref{2-9}.

\begin{figure} [tbh]
\epsfxsize=14cm
\hskip 17mm\mbox{\epsfbox{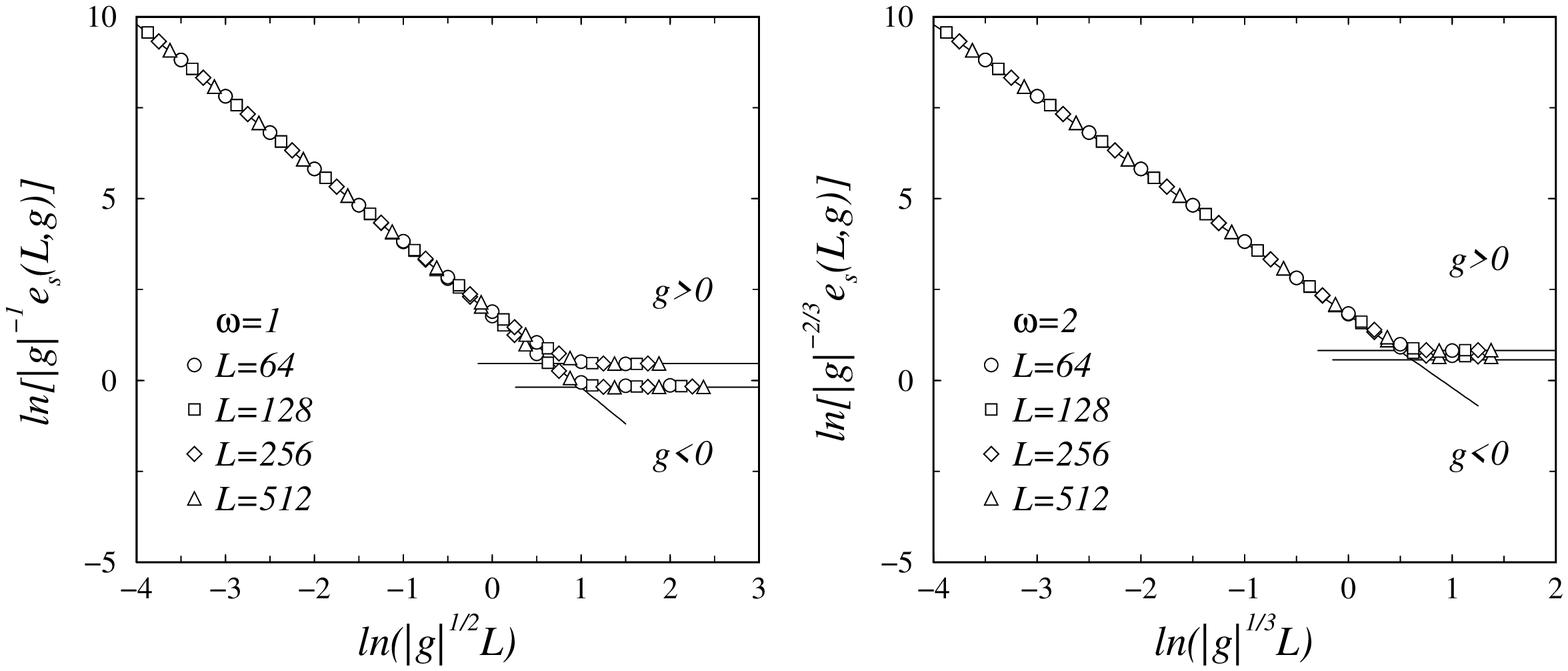}}
\vskip -.2cm
\caption{Finite-size scaling function of the surface energy density for $\omega=1$ (left) and $\omega=2$ (right). The lines indicate the slope -2 which follows from \eref{2-11} for $L\ll\ell$ and the constant asymptotic values for $L\gg\ell$.}
\label{fig7}  \vskip 0cm
\end{figure}

The surface energy density $e_{\rm s}$ is given by equation \eref{3-19} at $l=1$. Its finite-size behaviour is shown in figure \ref{fig7}. When $L\ll\ell$ the slope is equal to $-x_{\rm e_s}=-2$ as expected from \eref{2-11}. When $L\gg\ell$ the scaling function tends to a constant value which  is given by $g^{-1}e(z)$ at $z=1$ in the limit $g\to0$. For $\omega=1$ one obtains the constants $A_e^-$ in \eref{3-20} and $A_e^+$ in \eref{3-21}. In the logarithmic scale of figure \ref{fig7} this leads to $\ln A_e^-=-0.19112\dots$ and $\ln A_e^+=0.46736\dots$.

\section{Conclusion}
We have presented a theory for the scaling behaviour of physical profiles and finite-size behaviour in a semi-infinite ($z>0$) critical system  with free boundary condition in the presence of a space-dependent deviation from the critical coupling. The coupling is critical at the surface $z=0$ and deviates into the bulk as a power law, $\Delta(z)=g z^\omega$. The system is ordered (disordered) when the amplitude $g>0$ ($g<0$). The perturbation introduces a new length scale, which is given by $\ell \sim |g|^{-\nu/(1+\nu\omega)}$  and  depends on the correlation length exponent $\nu$ of the unperturbed system for a thermal-like perturbation.
This length scale appears as a supplementary relevant size which must be considered in the scaling theory. 

The scaling arguments have been confirmed by an exact solution of the semi-infinite Ising quantum chain with free boundary condition in the case of a linear variation of the transverse field and checked numerically for the quadratic case. In the linear case, in the scaling limit $L\rightarrow \infty$ and $g\rightarrow 0$ with $Lg$ fixed, the excitation spectrum of the Ising chain is exactly mapped onto  an harmonic oscillator problem with either Neumann or Dirichlet boundary conditions at the origin. Thus the magnetization and energy density profiles are given in terms of the harmonic oscillator eigenfunctions.

In this study we have considered the effect of gradient perturbations at the ordinary surface transition. Other types of surface transitions, such as special and extraordinary transitions, can be considered as well. We are currently investigating these issues.

\appendix
\setcounter{section}{1}
\section*{Appendix}
The surface magnetization in \eref{3-15} can be written as:
\be
m_{\rm s}=S^{-1/2}\,,\qquad S=1+\sum_{k=1}^{L-1}P_k^2\,,
\qquad \ln P_k=\sum_{l=1}^k\ln(1-gl^\omega)\,.
\label{a-1}
\ee
When $gL^\omega\ll1$ one obtains
\be
\ln P_k\simeq-g\sum_{l=1}^kl^\omega\simeq -\frac{gk^{\omega+1}}{\omega +1}\,,
\label{a-2}
\ee
for $k\gg1$. The sum in $S$ can be replaced by an integral when $L\gg1$, leading to:
\be
S\simeq\int_0^L{\rm d}k\,P_k^2\simeq\int_0^L{\rm d}k\,
\exp\left(-\frac{2gk^{\omega+1}}{\omega +1}\right)\,.
\label{a-3}
\ee

When $gL^{\omega+1}\ll1$ the exponential can be expanded and integrated term-by-term
\be
S\simeq L-\frac{2g}{\omega +1}\int_0^L{\rm d}k\,k^{\omega+1}
\simeq L\left(1-\frac{2gL^{\omega+1}}{(\omega +1)(\omega +2)}\right)\,,
\label{a-4}
\ee
and 
\be
m_{\rm s}\simeq L^{-1/2}\left(1+\frac{gL^{\omega+1}}{(\omega +1)(\omega +2)}\right)\,,
\label{a-5}
\ee
which gives equation \eref{3-17} when the scaling variable $u(L)$ is explicited.

When $gL^{\omega+1}\gg1$ and $g>0$, neglecting an exponentially small correction, one may rewrite \eref{a-3} as
\be\fl 
S\simeq\int_0^\infty{\rm d}k\, \exp\left(-\frac{2gk^{\omega+1}}{\omega +1}\right)
\simeq\frac{\int_0^\infty{\rm d}t\,t^{-\omega/(\omega+1)}\e^{-t}}{\left[2g(\omega+1)^\omega\right]^{1/(\omega+1)}}
\simeq\frac{\Gamma\left(\frac{1}{\omega+1}\right)}{\left[2g(\omega+1)^\omega\right]^{1/(\omega+1)}}\,,
\label{a-6}
\ee
which leads to the first expression of $m_{\rm s}$ in \eref{3-18}.

Finally let us consider the case where $gL^{\omega+1}\gg1$ and $g<0$. Then in equation ~\eref{a-3} the argument of the exponential can be replaced by its first-order expansion near $L$, where the integrand takes its maximum value on the interval of integration
\be\fl
f(k)\simeq f(L)+(k-L)f'(L)\,,\qquad f(L)=\frac{2|g|L^{\omega+1}}{\omega +1}\,,
\qquad f'(L)=2|g|L^{\omega}\,,
\label{a-7}
\ee
and 
\be
S\simeq \e^{f(L)}\int_0^L{\rm d}k\, \e^{(k-L)f'(L)}\simeq \frac{\e^{f(L)}}{f'(L)}\,,
\label{a-8}
\ee
where an exponentially small contribution from $k=0$ has been neglected.
Equations \eref{a-3}, \eref{a-7} and~\eref{a-8} lead to the second expression of $m_{\rm s}$ in \eref{3-18}.

\end{document}